\documentclass[sigconf]{acmart}

\AtBeginDocument{%
  \providecommand\BibTeX{{%
    \normalfont B\kern-0.5em{\scshape i\kern-0.25em b}\kern-0.8em\TeX}}}

\usepackage{pgf}
\usepackage{tikz}
\usepackage{colortbl}
\usepackage{enumitem}
\usepackage{multirow}
\usepackage{makecell}
\usepackage{amsthm}
\usepackage{amsfonts}
\usepackage{url}
\usepackage{subfigure}
\newcommand{\ie}{\emph{i.e., }}

\newcommand{\etc}{\emph{etc.}}
\newcommand{\wrt}{\emph{w.r.t. }}

\begin{document}

\title{Integrate Temporal Graph Learning into LLM-based Temporal Knowledge Graph Model}

\author{He Chang}
\affiliation{%
  \institution{Communication University of China}
  \country{}
}
\email{hechangcuc@cuc.edu.cn}


\author{Jie Wu}
\affiliation{%
  \institution{Communication University of China}
  \country{}
}
\email{wujie@cuc.edu.cn}

\author{Zhulin Tao}
\authornote{Corresponding Author}
\affiliation{%
  \institution{Communication University of China}
  \country{}
}
\email{taozhulin@gmail.com}

\author{Yunshan Ma}
\affiliation{
  \institution{Singapore Management University}
  \country{}
}
\email{ysma@smu.edu.sg}

\author{Xianglin Huang}
\affiliation{%
  \institution{Communication University of China}
  \country{}
}
\email{huangxl@cuc.edu.cn}

\author{Tat-Seng Chua}
\affiliation{
  \institution{National University of Singapore}
  \country{}
}
\email{dcscts@nus.edu.sg}


\begin{abstract}
Temporal Knowledge Graph Forecasting (TKGF) aims to predict future events based on the observed events in history. Recently, Large Language Models (LLMs) have exhibited remarkable capabilities, generating significant research interest in their application for reasoning over temporal knowledge graphs (TKGs). Existing LLM-based methods have integrated retrieved historical facts or static graph representations into LLMs. Despite the notable performance of LLM-based methods, they are limited by the insufficient modeling of temporal patterns and ineffective cross-modal alignment between graph and language, hindering the ability of LLMs to fully grasp the temporal and structural information in TKGs. To tackle these issues, we propose a novel framework TGL-LLM to integrate temporal graph learning into LLM-based temporal knowledge graph model. Specifically, we introduce temporal graph learning to capture the temporal and relational patterns and obtain the historical graph embedding. Furthermore, we design a hybrid graph tokenization to sufficiently model the temporal patterns within LLMs. To achieve better alignment between graph and language, we employ a two-stage training paradigm to finetune LLMs on high-quality and diverse data, thereby resulting in better performance.  
Extensive experiments on three real-world datasets show that our approach outperforms a range of state-of-the-art (SOTA) methods.
\end{abstract}





\maketitle

\section{Introduction}
\begin{figure}
    \centering
    \includegraphics[width=\linewidth]{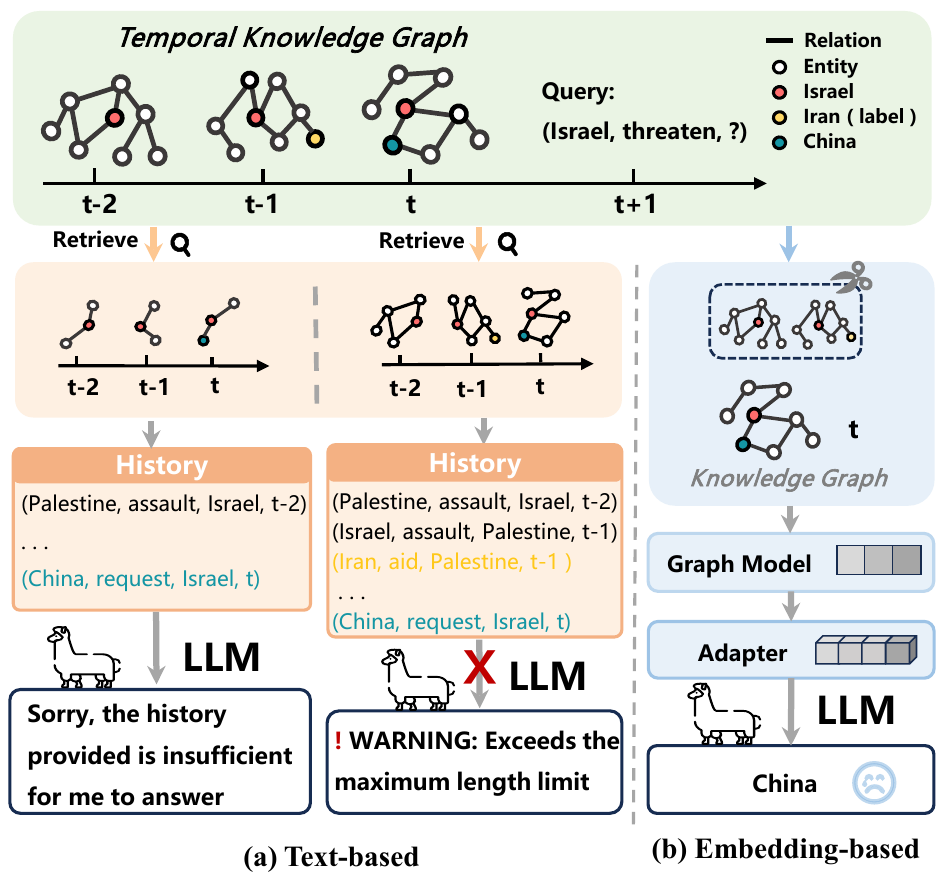}
    \caption{Illustration of the limitations in the current LLM-based TKGF methods. (a) Text-based: the cutting-edge retrieval-augmented solutions. (b) Embedding-based: incorporating the static knowledge graph into LLM.}
    \label{fig: introduction}
    \vspace{-6mm}
\end{figure}
Temporal knowledge graph forecasting (TKGF)~\cite{survey-of-tkg} aims to predict future facts by leveraging historical information, which plays a vital role in enhancing decision-making, resource allocation, and strategic planning across various domains~\cite{survey-event-forecasting}. 
Facts in temporal knowledge graphs (TKGs) represent real-world knowledge in a structured format, typically represented as a quadruple $(s, r, o, t)$, where $s$, $r$, $o$, and $t$ denote the subject, relation, object, and timestamp, respectively.
Previous research~\cite{REGCN,RENET,HiSMatch,SeCoGD} typically focuses on employing graph convolution networks (GCN) and recurrent neural networks (RNN) as the encoders to model the relational and temporal patterns hidden in historical facts. 
Despite their effectiveness, there are several limitations. 
First, these temporal graph neural network(TGNN) methods often fall short in being overly reliant on the training data, where the scope of knowledge is restricted when facing open-world data. 
Moreover, the errors induced by automatic information extraction systems bring substantial noise in TKGs~\cite{forecasttkgquestions}, potentially resulting in suboptimal graph representation learning.
Finally, the prevailing existence of long-tail entities~\cite{tackling-long-tail} with only a few observed links presents further challenges to the TGNN methods, which mostly follow the supervised learning paradigm and difficult to handle the long-tail issues. 

Scrutinizing these limitations, the emerging large language models (LLMs) ~\cite{GPT, RLHF, LLaMA} demonstrate great potential and have been explored in some pioneering works ~\cite{PPT, GPT-NeoX-ICL, GENTKG, Chain-of-History} on TKGF.
These works can be broadly categorized into two branches: text-based and embedding-based.
The text-based methods design a variety of retrieval strategies, such as temporal logical rules~\cite{GENTKG}, semantic similarity~\cite{MM-Forecast}, and certain heuristics~\cite{Chain-of-History}, to transform the graph data into textual format for LLMs' reasoning.
However, as illustrated in Figure~\ref{fig: introduction}, the varying retrieval strategies may inadvertently omit critical contextual details (\ie \textit{(Iran, aid, Palestine, t-1)}), and the limited token length makes it impractical to incorporate all the historical contextual information into LLMs. 
To bridge the gap between structured graph data and natural language, researchers propose the embedding-based methods~\cite{graphtranslator, KoPA} to encode the graphs into graph representations with a pre-trained graph model, followed by an adapter to align the graph representations with the language model. 
Compared to text-based methods, the aligned graph representation incorporates structural information into LLMs, providing valuable historical context for TKGF. However, existing embedding-based methods still face two unresolved limitations:
\begin{itemize}[leftmargin=*]
\item  \textbf{\textit{Insufficient modeling of temporal patterns.}}
Existing embedding-based methods~\cite{graphtranslator, KoPA} primarily focus on reasoning over static knowledge graphs, which fails to fully exploit the temporal reasoning ability of LLM. 
For instance, as illustrated in Figure~\ref{fig: introduction}, due to the absence of temporal information, LLMs provide the incorrect answer '\textit{China}' based on the historical facts at timestamp $t$. Therefore, the sufficient modeling of temporal patterns in LLMs remains an area that requires further exploration.
\item  \textbf{\textit{Ineffective cross-modal alignment between graph and language.}}
The inherent noise and long-tail issues present in TKGs result in suboptimal graph representations, which becomes increasingly challenging for the cross-modal alignment between graph and language. Furthermore, large-scale real-world knowledge exhibits a wide range of temporal and relational patterns, while only fine-tuning LLMs with similar patterns may degrade the performance in complex scenarios. Consequently, it is crucial to automatically identify high-quality and diverse graph data from the vast ocean of available datasets.

\end{itemize}

In order to advance the embedding-based approach by addressing the aforementioned limitations, we propose a novel LLM-based framework for TKGF, termed \textbf{TGL-LLM}, which integrates temporal graph learning into LLMs. 
To fully exploit the temporal reasoning ability of LLMs, we first introduce temporal graph learning to capture temporal and relational patterns. Rather than directly using the final representation in temporal graph learning, we extract recent historical graph embedding through GCNs to preserve temporal information of the evolving facts.
More importantly, we design a hybrid graph tokenization to sufficiently model the temporal patterns in TKGs within LLMs.
Specifically, we introduce a trainable temporal graph adapter to project the graph embedding into the language token space and concatenate the graph tokens in temporal order as the description of the entity, thereby allowing LLMs to further explore complex temporal patterns.
Additionally, to achieve effective cross-modal alignment between graphs and language, we employ a two-stage fine-tuning paradigm to fine-tune LLMs on high-quality and diverse data, resulting in the improvement of performance. To identify the high-quality and diverse data, we first introduce the influence function to evaluate the influence of removing each graph data point on the fine-tuning of LLM. A higher influence score indicates that the graph data is of higher quality for the LLM to effectively capture the temporal and relational patterns in graph embedding.
Then we apply diversity sampling to ensure a variety of temporal and relation patterns exposed to LLMs.
Extensive experiments demonstrate that our method outperforms both traditional approaches and state-of-the-art (SOTA) methods. In summary, our contributions can be outlined as follows:
\begin{itemize}[leftmargin=*]
    \item We advance embedding-based LLM solutions for TKGF by addressing two limitations, \ie insufficient modeling of temporal patterns and ineffective cross-modal alignment between graph and language.
    \item We propose a novel framework TGL-LLM, where the two key designs, hybrid graph tokenization and a two-stage training paradigm, pertinently alleviating the two limitations respectively.
    \item We conduct extensive experiments on three real-world datasets, and the results demonstrate the effectiveness of TGL-LLM.
\end{itemize}
\section{Related Work}
We provide a brief literature review of methods for TKGF, including both conventional Non-LLM methods and LLM-based methods. 
\subsection{Temporal Knowledge Graph Forecasting}
Temporal Knowledge Graph Forecasting is a rapidly evolving field that focuses on predicting the occurrence of future facts based on historical data. Traditional methods can generally be grouped into two categories: rule-based methods and GNN-based methods. Rule-based methods~\cite{tlogic, bai2023temporal} rely on predefined or learned temporal rules that capture common patterns or logical relationships in historical event data to guide the forecasting process. For instance, utilizes temporal random walks in TKGs to learn temporal logical rules without the need for trainable parameters, thereby enhancing the explainability of the reasoning process. TILP ~\cite{TILP} proposes a differentiable framework for temporal logical rule learning by introducing a constrained walk mechanism and temporal operators, enabling symbolic learning in temporal knowledge graphs.  TEILP~\cite{Teilp} extends the framework of TILP to address the challenge of time prediction by constructing a temporal knowledge graph and developing a differentiable random walk approach for time prediction. However, these methods depend heavily on manually defined rules, which involve a significant time investment when applied to large-scale TKG datasets. On the other hand, TGNN-based methods~\cite{REGCN,RENET} employ graph neural networks to aggregate relational information from entity neighborhoods.  Early works such as RGCN~\cite{RGCN} and ConvTransE~\cite{ConvTransE} treat event forecasting as a static knowledge graph completion task, updating entity and relation embeddings through GNNs.  To capture the temporal evolution of entities and relations, methods like RENET and REGCN incorporate recurrent neural networks to derive temporal representations. Addressing the absence of specific and informative facts, SeCoGD~\cite{SeCoGD} uses textual topic modeling to disentangle subgraphs, while LoGo~\cite{LoGo} applies text clustering to construct complex facts for forecasting using both local and global contexts.
To incorporate textual information into representation learning, Glean~\cite{Glean} and CMF~\cite{CMF} fuse textual embeddings into graph edges. However, due to the limitations inherent in GNNs, these methods suffer from the inherent noise and long-tail entities in TKGs.

\subsection{Large Language Models for TKGF}
LLMs have recently been explored for event forecasting, leveraging their ability to process and understand contextual information. Several studies have highlighted the forecasting potential of in-context learning within large language models. Following the retrieval-augmented generation paradigm, several recent works employ different sampling strategies, flattening the structural graph data into textual descriptions to predict future facts. For instance, GPT-NeoX-ICL~\cite{GPT-NeoX-ICL} retrieves relevant facts from the temporal knowledge graph and constructs prompt with a list of historical facts in quadruplet format. By utilizing an in-context learning framework, LLMs can comprehend the task of TKGF and the reason for the given historical information. Similar to TLogic~\cite{tlogic}, GENTKG~\cite{GENTKG} utilizes learned temporal logical rules to construct the relevant event histories. CoH~\cite{Chain-of-History} retrieves the results based on the well-designed rules to guide LLMs in making the final prediction. However, the limitations of the natural language format result in insufficient contextual information for effective reasoning on the temporal knowledge graph. In contrast to transforming graph data into textual descriptions, recent studies treat pre-trained graph embeddings as a new modality for LLMs, aligning them with language space. GraphTranslator~\cite{graphtranslator} generates alignment data to aid the translator module in converting graph embeddings to token embeddings. TEA-LLM~\cite{TEA-GLM} aligns GNN representation with LLM token embeddings by a linear projector to make LLMs understand the structural information in various tasks.  KoPA~\cite{KoPA} introduces an adapter to obtain aligned structural embeddings, positioning them as a prefix to the input prompt. However, existing alignment methods assume pre-trained graph embeddings as ideal representations of structural information, overlooking the suboptimal representations caused by noise and long-tail entities in the training data. More importantly, all of the above embedding-based methods focus on static knowledge graph reasoning, neglecting the temporal dimension essential for event forecasting. 
\section{Preliminary}
\label{preliminary}
We first present the classicial formulation of TKGF. Then we present its adaption in the scenario of LLM-based solutions.

\noindent\textbf{Classical TKGF Formulation.} 
Given a set of entities $\mathcal{E}$ and relations $\mathcal{R}$, we define a fact as a quadruple $(s, r, o, t)$, where $s\in\mathcal{E}$, $r\in \mathcal{R}$, and $o\in \mathcal{E}$ represent the subject, relation, object, and timestamp respectively. All facts occurring at the same timestamp $t$ form a knowledge graph denoted as $\mathcal{G}_t=\{(s_n, r_n, o_n, t)\}_{n=1}^{N}$, where $(s_n, r_n, o_n, t)$ is the n-th fact and $N=|\mathcal{G}_t|$ is the number of facts at timestamp $t$. Consequently, a TKG $\mathcal{G}$ can be formalized as $\mathcal{G} = \left \{ \mathcal{G}_0, \mathcal{G}_1, \ldots, \mathcal{G}_{t}, \ldots \right \}$. Given the historical knowledge graphs prior to $t$, denoted as $\mathcal{G}_{\le t} = \left \{ \mathcal{G}_0, \mathcal{G}_1, \ldots, \mathcal{G}_{t} \right\}$, and a query of the form $(s, r, ?, t)$, temporal knowledge graph forecasting aims to predict the missing object.

\noindent\textbf{LLM-based TKGF Formulation.} Compared to traditional methods that rank all entities within the graph, LLM-based methods face practical limitations due to the large size of the entity set and the constrained token length of LLMs, making it infeasible to input all entities into the model simultaneously. Therefore, we simplify the temporal knowledge graph forecasting settings as a Multi-Choice Questions(MCQ) problem, employing negative sampling strategies to construct a candidate set $\hat{O}=\left \{\hat{o}_{1}, \hat{o}_{2}, \ldots, \hat{o}_{k}, \ldots\right\}$ for each query, where $K=|\hat{O}|$ denotes the number of candidate entities. The reasons are as follows: First, it facilitates easier and more consistent quantitative evaluation for LLM-based methods. Second, the MCQ setting aligns closely with real-world scenarios since there are typically a limited number of highly probable candidate entities rather than an unrestricted range of possibilities. Third, even in a generative setting, the generated answers are more likely to be derived from the provided context rather than from independent reasoning based on historical information~\cite{zero-shot-rankers}. For more details, refer to Appendix ~\ref{statistic}. Consequently, we argue that the MCQ setting is a reasonable and acceptable choice for this task. Notably, the MCQ setting differs from the typical Question and Answer(QA) task, which emphasizes factual retrieval and reasoning based on the current context, typically lacking the capacity to infer dynamic changes over time. Given the historical knowledge graphs prior to $t$, denoted as $\mathcal{G}_{\le t} = \left \{ \mathcal{G}_0, \mathcal{G}_1, \ldots, \mathcal{G}_{t} \right\}$, a query of the form $(s, r, ?, t)$, and a candidate set $\hat{O}$, temporal knowledge graph forecasting aims to predict the missing object from the candidate set.

\section{Method}
\begin{figure*}
    \centering
    \includegraphics[width=\linewidth]{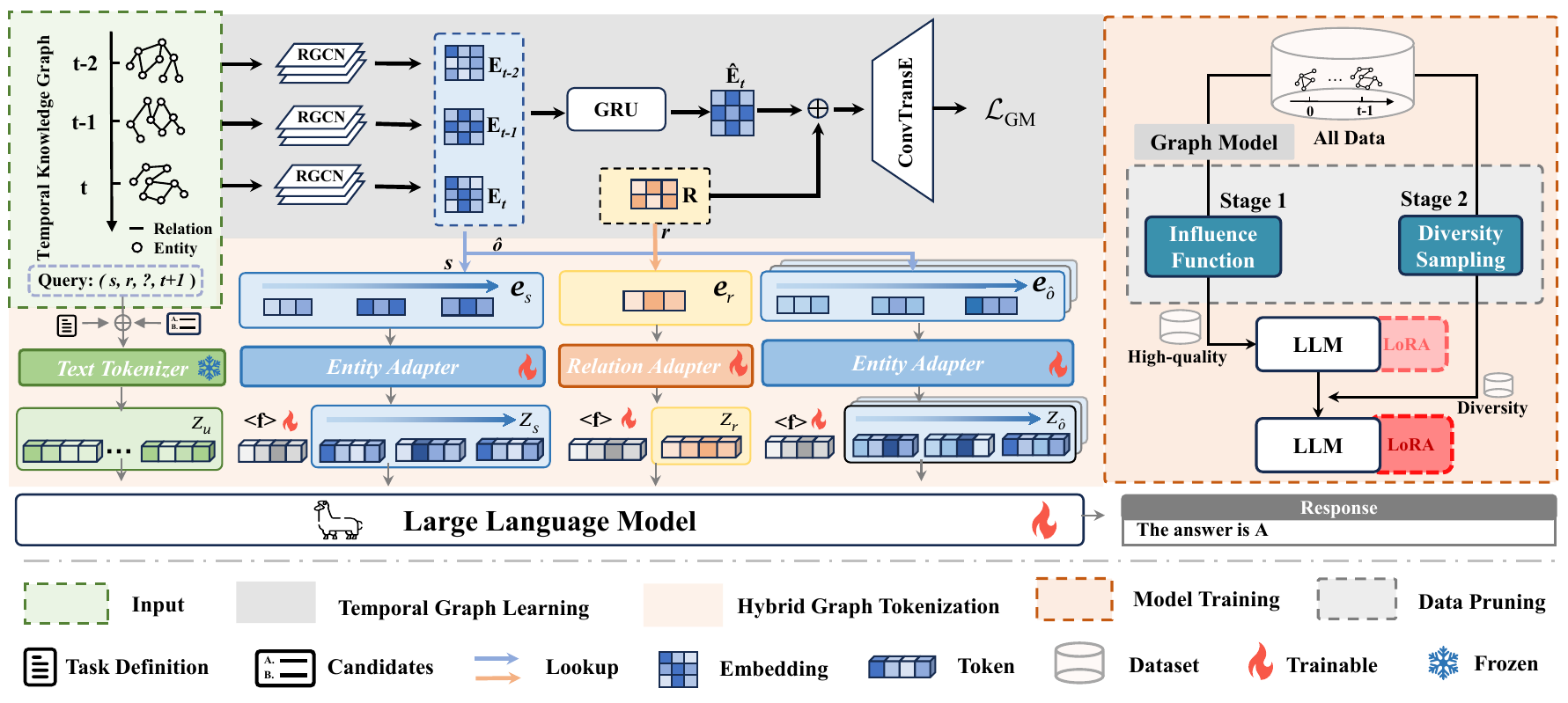}
    \caption{The overall framework of TGL-LLM consists of two key novel modules: 1) hybrid graph tokenization that sufficiently models the temporal patterns in LLMs (left), and 2) model training that follows a two-stage training paradigm (right).}
    \label{fig: framework}
    \vspace{-3mm}
\end{figure*}

We present a novel framework TGL-LLM to integrate temporal graph learning into LLM-based temporal knowledge graph model, as illustrated in Figure ~\ref{fig: framework}. At the core of TGL-LLM are two key innovations: 1) a novel hybrid graph tokenization, to sufficiently model the temporal patterns, and 2) a two-stage training paradigm, to achieve effective cross-modal alignment between graph and language. In this section, we first present temporal graph learning, where we capture the temporal and relational patterns and obtain the temporal graph embedding. Then, we introduce hybrid graph tokenization to enable LLMs to capture the complex temporal patterns in TKGs. Finally, we introduce model training, which employs a data pruning method to identify high-quality graph data and diverse graph data, followed by a two-stage training paradigm.

\subsection{Temporal Graph Learning}
The representation learning of temporal graph has been continuously under development. To sufficiently model the temporal patterns, we introduce temporal graph learning to obtain the historical graph embedding. 
To efficiently leverage the temporal reasoning ability of LLMs, we utilize the historical graph embedding instead of the final representation to enable LLMs to preserve temporal information of the evolving facts. Meanwhile, in the training phase, we train the graph model with an RNN and a decoder to capture the temporal patterns within the recent historical graph representation.

Following the previous work REGCN~\cite{REGCN}, we employ RGCN as the graph model to capture the relational patterns among all facts at the same timestamp. Based on the $\mathcal{G}_{t}$ at timestamp $t$, for each layer $l$ of the graph propagation, the message $e_{o}^{l} \in \mathbb{R}^{d}$ obtained by each object $o$ is defined as follows:
\begin{equation}
\mathbf{e}_{o}^{l}=f(\frac{1}{\left |  \mathcal{G}_{t} \right | } \sum_{(s,r,o)\in\mathcal{G}_{t}} \mathbf{W}_{1}^{l}(\mathbf{e}_{s}^{l-1}+\mathbf{r})+\mathbf{W}_{2}^{l}\mathbf{e}_{o}^{l-1}),
\end{equation}
where $\mathbf{W}_{1}^{l}, \mathbf{W}_{2}^{l} \in \mathbb{R}^{d}$ are the parameters of convolutional kernel in the $l-1$-th layer, $d$ is the dimension of the representation, $f(\cdot)$ is the activation function RReLU, and $\mathbf{e}_{s}^{l-1}$, $\mathbf{e}_{o}^{l-1}$, and $\mathbf{r}$ denote the $l$-th embeddings of entities $o$, $s$ and  relation $r$. After executing multi-layer message passing, we aggregate the messages acquired from multi-layer propagation to derive the entity representation at timestamp $t$, formally defined as:
\begin{equation}
\mathbf{e}_{o} = \sum_{l=0}^{L}\mathbf{e}_{o}^{l},
\end{equation}
where $L$ is the number of graph layers. Through the message-passing mechanism, the embedding for all entities at timestamp $t$ and all relations are denoted as $\mathbf{E}_{t} \in \mathbb{R}^{|\mathcal{E}|*d}$ and $\mathbf{R} \in \mathbb{R}^{|\mathcal{R}|*d}$, respectively. It is noted that we adopt the recent T steps embedding of entities $\mathbf{E}=\{\mathbf{E}_{t-T} ... \mathbf{E}_{t}\}$ and relations $\mathbf{R}$ as our graph representaion.

In order to capture the temporal evolving patterns in TKG, in the training phase, a GRU is employed to capture the temporal patterns, defined as follows: 
\begin{equation}
\hat{\mathbf{E}}_{t} = \text{GRU}(\mathbf{E}_{t},\mathbf{E}_{t-1}).
\end{equation}
Subsequently, we utilize ConvTransE~\cite{ConvTransE} as the decoder and apply the cross-entropy loss to optimize the model. The loss is defined as:
\begin{equation}
\mathcal{L}_{GM}=\sum_{(s,r)\in \mathcal{G} _{t+1}}y_{(s,r,t+1)}\text{log}(p(\mathcal{E}|s,r,\mathcal{G}_{\le t})),
\end{equation}
where $y$ denotes the ground truth. In this way, the pre-trained graph model can extract temporal information from TKG, thereby enhancing the capability of LLMs for more profound temporal reasoning.

\subsection{Hybrid Graph Tokenization}
\subsubsection{Temporal Graph Adapter}
To bridge the gap between graph and language, directly incorporating graph embedding into the prompt can not be interpreted by LLMs. Therefore, following previous embedding-based methods~\cite{graphtranslator,TEA-GLM}, we first map the graph embedding into the language token space, allowing LLMs to understand the temporal and structural information distilled by the pre-trained graph model. To achieve this, we introduce a trainable Temporal Graph Adapter (TGA): Entity Adapter(EA) and Relation Adapter(RA).

Given the most recent T-step graph embedding of entity $\mathbf{E}$ and relation representation $\mathbf{R}$, we can obtain the temporal graph representation for the subject $e_{s}=\left \{e_{t-T}^{s},\ldots,e_{t}^{s}\right\}$ and the candidate $e_{\hat{o}_{k}} = \left \{e_{t-T}^{\hat{o}_{k}},\ldots,e_{t}^{\hat{o}_{k}}\right\}$, along with relation representation $r$. Then we project the temporal graph representation into the language token space with the Entity Adapter and Relation Adapter:
\begin{align}
    & z_{t-T}^{s}=\text{EA}(e_{t-T}^{s}; \Theta_{ea}) \\
    & z_{r} = \text{RA}(r; \Theta_{ra}) \\
    & z_{t-T}^{\hat{o}_{k}} = \text{EA}(e_{t-T}^{\hat{o}_{k}}; \Theta_{ea}),
\end{align}
where $z_{t-T}^{s}$ and $z_{t-T}^{\hat{o}_{k}}$ denote the temporal graph tokens of subject and candidate at timestamp $t-T$, respectively; $z_{r}$ is the graph token of relation; $\text{EA}\left ( \cdot  \right ) $ and $\text{RA}\left ( \cdot  \right ) $ denote the entity adapter and relation adapter; $\Theta_{ea}$ and $\Theta_{ra}$ are the parameters of entity adapter and relation adapter. Here, we employ a two-layer perception to serve as the adapter of entity and relation.

\subsubsection{Hybrid Prompt Design}
To help LLMs understand and capture complex temporal patterns, we propose a hybrid prompting method. Specifically, the designed hybrid prompt consists of three primary components:

\noindent\textbf{Instruction.} 
We first provide a clear task definition of TKGF. Then we give the query fact and the candidate set in natural language format. The instruction, which consists of task definition, query, and the candidate set, is then transformed into tokens represented as $z_{u}$. Figure~\ref{fig:instruction} in Appendix~\ref{instruction} illustrates the instructions we utilized.

\noindent\textbf{Query with Graph Tokens.} We concatenate the temporal graph tokens in temporal order as the description of the subject, denoted as $z_{s}$. Furthermore, we introduce a learned feature token $<\mathrm{f}>$ as the soft prompt to distinguish between textual tokens and various graph tokens, denoted as:
\begin{equation}
    z_{q}=[\underbrace{z_{f}}_{<\mathrm{f} >} , \underbrace{z_{t-T}^{s},\ldots,z_{t}^{s}}_{\mathrm{subject}} ],
\end{equation}
where $z_{q}$ is the query subject with graph token and $z_{f}$ denotes the learned feature token. Subsequently, The graph token of the query relation $z_{r}$ is subsequently concatenated with $z_{f}$ to form the description of the relation.

\noindent\textbf{Candidate Set with Graph Tokens.} Similar to the query subject, we concatenate the feature token $z_{f}$ and temporal graph tokens $z_{\hat{o}_{k}}$ to construct the description of candidates $z_{\hat{o}}$. 

Notably, the hybrid prompt only utilizes temporal graph tokens as the description of the subject and the candidate, which facilitates the integration of temporal graph representation sourced from temporal graph learning. This design addresses the limitations of prompts that rely either on textual historical facts or static graph representation, thereby generating more accurate predictions.

\subsection{Model Training}
\subsubsection{Data Pruning}
To achieve more effective cross-modal alignment between the graph and the language, the key is the selection of high-quality, diverse data. Suboptimal graph representations caused by the noise and long-tail issuse in TKGs increase the difficulty of alignment, while redundant samples with similar patterns restrict the capacity of LLMs to learn a wide range of domain knowledge. Therefore, first, we introduce the influence function~\cite{IF} as the data pruning method to evaluate the influence of each training sample in TKGs. 

Considering the heavy computational resources and time costs of LLMs, following previous studies~\cite{data_efficient}, we utilize the pre-trained graph model as the surrogate model to evaluate the influence score. Given the parameters of the pre-trained graph model $\theta_{G} \in \Theta_{G}$, the train dataset $\mathcal{D}$, and a training sample $d$ upweighted by a small $\epsilon$, the empirical risk minimizer is given by:
\begin{equation}
    \hat{\theta}_{G} = \underset{\theta_{G}\in \Theta_{G}}{\mathrm{argmin}} \frac{1}{m} \sum_{i=1}^{m}\mathcal{L} (d_{i},\theta_{G})+\epsilon \mathcal{L}(d,\theta_{G}),  
\end{equation}
where $m=|\mathcal{D}|$ is the number of train dataset and $d_{i}\in\mathcal{D}$ is the training sample. Notably, removing a sample $d$ is the same as assigning $-\frac{1}{m}$ to $\epsilon$. Based on the classic result ~\cite{ling1984residuals}and the chain rule, the influence of removing a sample $d$ on the loss of an arbitrary sample $d'$ is given by:
\begin{equation}
\begin{split}
  \mathcal{I}_{\mathrm{remove,loss} }(d,d') &= \left.\frac{\mathrm{d}\mathcal{L}(d',\hat{\theta}_{G})}{\mathrm{d} \epsilon }\right |_{\epsilon =0} \\ 
&=\nabla_{\theta_{G}}\mathcal{L}(d',\hat{\theta}_{G} )^{T} \left.\frac{\mathrm{d}\hat{\theta }_{G}}{\mathrm{d}\epsilon }\right |_{\epsilon =0}\\
&=\frac{1}{m} \nabla_{\theta_{G}}\mathcal{L}(d',\hat{\theta}_{G} )^{T}H_{\hat{\theta}_{G}}^{-1}\nabla_{\theta_{G}}\mathcal{L}(d,\hat{\theta}_{G}),
\end{split}
\end{equation}
where $H_{\hat{\theta}_{G}}^{-1} = \frac{1}{m}\sum_{i=1}^{m}\nabla_{\theta_{G}}^{2}\mathcal{L} (d_{i},\theta_{G})$ is the Hessian and positive definite by assumption. Furthermore, we utilized the Hessian-vector products (HVP) ~\cite{agarwal2016second} and symmetric property to obtain influence scores for each $d_{i}\in\mathcal{D}$ in train data:
\begin{equation}
     \mathcal{I}_{\mathrm{remove,loss} }(d,\mathcal{D}) = \frac{1}{m} \nabla_{\theta_{G}}\mathcal{L}(d,\hat{\theta}_{G})^{T} H_{\hat{\theta}_{G}}^{-1} \left (\sum_{i=1}^{m} \frac{1}{m} \nabla_{\theta_{G}}\mathcal{L}(d_{i},\hat{\theta}_{G} )\right ).
\end{equation}
Benefiting from the auto-grad system, we only need to specify the loss to compute the influence score. Subsequently, we apply stratified sampling~\cite{coverage-sampling} to obtain a high-quality subset $\mathcal{D}_{h}$, ensuring a reduction in similar training samples.

In addition to quality, the effectiveness of LLM-based methods fundamentally depends on the diversity of training data~\cite{miranda2023beyond}, which enables LLMs to capture various temporal and relational patterns in TKGs effectively. In other words, diverse data is essential for instilling the general reasoning capabilities of LLMs for TKGF. Therefore, we selected a subset $\mathcal{D}_{p}$ from the original dataset as the implementation of the high-quality subset, termed the diversity subset. Here, we employ random sampling for simplicity and efficiency to obtain the diversity subset. It is important to note that, to avoid impacting the representation learning on high-quality data, the size of the diversity subset is significantly smaller than that of the high-quality subset.

\subsubsection{Prompt Tuning}
Based on the aforementioned subset of train data, we train the model following a two-stage training paradigm. Compared to the random sample data, the two-stage training phase on the high-quality subset and diversity subset of train data better aligns the temporal and global graph representations with the token space and enhances the robustness of LLMs in TKGF.

\noindent\textbf{Stage 1.} Given the high-quality subset $D_{h}$ with the designed hybrid prompt, we adopt LoRA for the finetuning of LLM. The optimizing objective of LoRA can be formulated as follows:
\begin{equation}
    \max_{\Theta  } \sum_{d\in \mathcal{D}_{h} } \text{log}(p  (v_{o}|z_{u},z_{q},z_{\hat{o}};\Phi_{0}, \Phi(\Theta_{LoRA}),\Theta_{ea}, \Theta_{ra}, \Theta{f}),
\end{equation}
where $\Phi_{0}$ and $\Phi(\Theta_{LoRA})$ represent the frozen parameters of LLMs and the learnable parameters of LoRA, respectively; $d\in \mathcal{D}_{h}$ denote the training sample in the high-quality subset.

\noindent\textbf{Stage 2.} Based on the fine-tuned LLM on the high-quality subset, we aim to improve the generality reasoning ability of LLM. To this end, analogous to the training phase of stage 1, we continue to fine-tune the LLM on the diversity subset $\mathcal{D}_{p}$ with the above parameters $\Phi(\Theta_{LoRA})$, $\Theta_{ea}$, $\Theta_{ra}$, $\Theta{f}$.
\section{Experiments}
We conduct experiments on three real-world datasets and compare the results of our method with several baselines, including conventional Non-LLM methods and LLM-based approaches. Additionally, we perform ablation studies to demonstrate the significant improvements resulting from the integration of temporal and global graph representations, as well as data pruning. Furthermore, we examine the effects of historical length, and the influence on long-tail entities, and present case studies to clearly illustrate our advantages over the baselines. To validate the superiority of our framework, we aim to answer the following research questions:
\begin{itemize}[leftmargin=*]
    \item \textbf{RQ1}: How does TGL-LLM perform compared to conventional TKGF models and LLM-based methods?
    \item \textbf{RQ2}: How does the integration of temporal and global graph representations affect performance? Are our data pruning methods superior and beneficial?
    \item \textbf{RQ3}: How does TGL-LLM perform under different settings, including varying historical lengths and long-tail entities?
\end{itemize}

\subsection{Experiment Settings}
\subsubsection{Datasets}
\label{datasets}

We utilize the recently proposed large-scale POLECAT dataset and crop three subsets to conduct experiments, \ie POLECAT-IR, POLECAT-IS, and POLECAT-EG. POLECAT is a temporal knowledge graph dataset in socio-political domain. Even though there are several widely-used TKG datasets, such as ICEWS \cite{icews}, GDELT \cite{gdelt}, WIKI \cite{wiki}, and YAGO \cite{yago}, they were constructed quite long time ago, consequently, the knowledge in these datasets may have unintentionally leaked into the pre-training data of LLMs \cite{halawi2024approaching}. The training data leakage raise the concerns of unfair or unfaithful evaluation of the TKGF methods, because we cannot tell the performance improvement comes from better forecasting capability of the designed model or just because the LLMs already know the future facts. Although prior efforts have tried to mitigate leakage by anonymizing timestamps \cite{lee2023temporal}, LLMs can still recognize the semantics of entity names and event types, linking them to internal knowledge. To ultimately eliminate the data leakage concern, we adopted the POLECAT dataset \cite{polecat}, which was first released in April 2023 as an upgraded version of ICEWS. The biggest merit is that this is a live dataset which keeps updating new knowledge every week. And it introduces an upgraded schema, utilizes an enhanced information extraction pipeline with less noise, and provides a large-scale dataset enriched with various metadata, including location, context, \etc \ We use Llama2 as our default LLM backbone, and the training data cut-off date of Llama2 is July 2023. Regarding this cut-off date, we split the dataset along the timeline into training (January 2018 to May 2023), validation (June to October 2023), and test (November 2023 to April 2024), assuring all the test data and a part of the validation data are after the LLM cut-off date. Given the large scale of the original POLECAT dataset, we follow the previous approaches \cite{deng2020dynamic,SeCoGD} and crop three sub-datesets according to the country labels of each quadruple, \ie Iran, Israel, and Egypt, corresponding to POLECAT-IR, POLECAT-IS, and POLECAT-EG, respectively. Table \ref{tab:subdatasets_stats} presents the statistics of the datasets, where $|\mathcal{R}|$ denotes the number of relations and $|\mathcal{E}|$ denotes the total number of entities, showing that they are large-scale in terms of the entity set and the total number of quadruples. And we also present the detailed pre-processing steps in Appendix \ref{dataset_constr} and extensive statistical analysis in Appendix \ref{additional_stats}.

\begin{table}[h]
    \centering
    \vspace{-3mm}
    \caption{Dataset statistics, including the number of facts, entities $|\mathcal{E}|$, and unique relations $|\mathcal{R}|$ for training, validation, and testing sets in each of the dataset.}
    \vspace{-3mm}
    \renewcommand{\arraystretch}{1.2}
    \resizebox{\linewidth}{!}{
    \begin{tabular}{l|ccc|ccc|ccc}
        \toprule
        \multirow{2}{*}{\textbf{Set}}& \multicolumn{3}{c|}{\textbf{POLECAT-IR}} & \multicolumn{3}{c|}{\textbf{POLECAT-IS}} & \multicolumn{3}{c}{\textbf{POLECAT-EG}} \\
         & \textbf{Facts} & \textbf{$|\mathcal{E}|$} & \textbf{$|\mathcal{R}|$} & \textbf{Facts} & \textbf{$|\mathcal{E}|$} & \textbf{$|\mathcal{R}|$} & \textbf{Facts} & \textbf{$|\mathcal{E}|$} & \textbf{$|\mathcal{R}|$} \\
        \midrule
        \textbf{Train} & 616,880 & 33,920 & 80 & 484,630 & 31,898 & 80 & 256,523 & 25,422 & 80 \\
        \textbf{Valid} & 13,737 & 2,593 & 79 & 29,931 & 4,797 & 77 & 6,898 & 1,399 & 72 \\
        \textbf{Test} & 11,053 & 2,174 & 72 & 56,750 & 6,364 & 77 & 3,812 & 1,096 & 69 \\
        \midrule
        \textbf{Total} & 641,670 & 35,138 & 80 & 571,311 & 36,336 & 80 & 267,233 & 26,220 & 80 \\
        \bottomrule
    \end{tabular}}
    \label{tab:subdatasets_stats}
 \vspace{-3mm}

\end{table}

\subsubsection{Evaluation Protocols}
For each query event, we construct the candidate set by randomly selecting 4, 6, or 10 entities from the most recent three days, including one positive entity. We employ \textit{Accuracy} as the primary metric to evaluate the ability of models to predict the positive item from this candidate set. A higher Acc@4, Acc@6, or Acc@10 indicates a superior capability for accurate temporal knowledge graph forecasting. To ensure a fair comparison across all baselines, we utilize the same number of candidates and evaluation methods.

\subsubsection{Baselines}
We compare our method with three types of baseline methods, \ie Non-LLM methods, and LLM-based methods including zero-shot and fine-tune approaches, which are widely used in the TKG formulation. 

\noindent\textbf{Non-LLM Methods} convert TKG into embeddings, including DistMult~\cite{DistMult}, ConvTransE~\cite{ConvTransE}, RGCN~\cite{RGCN}, RENET~\cite{RENET}, REGCN~\cite{REGCN}, and HisMatch~\cite{HiSMatch}.

\noindent\textbf{Zero-Shot Methods} use TKG information in text form in the LLMs prompt. We evaluate results using GPT-3.5-turbo ~\cite{OpenAI2022ChatGPT} and GPT-4o-mini~\cite{achiam2023gpt}.

\noindent\textbf{Fine-tune Methods} use TKG to train LLMs, including pure text form of TKG data (e.g., GenTKG~\cite{GENTKG}, CoH~\cite{Chain-of-History}), and direct use of graph data as input (e.g., KoPA~\cite{KoPA}).

\subsubsection{Implementation Details}
To ensure fairness of evaluation, all conventional Non-LLM methods employ the Adam optimizer, with the learning rate selected from \{1e-2, 1e-3, 1e-4\} and L2 regularization from \{1e-4, 1e-5, 1e-6\}. The embedding size is set to 200, and the number of propagation layers is set to 2. For all LLM-based methods, we utilize Llama2-7B-chat~\cite{llama2} as the backbone, and each experiment is trained on 100,000 training samples randomly selected from the training dataset, with a batch size of 128. Moreover, the sizes of the high-quality subset and diversity subset we utilized are set to 100,000 and 10,000, respectively. For the embedding-based method, the graph embedding size is set to 200. Notably, based on the split of the dataset, we adopt the real timestamp in all the LLM-based method instead of the index. All experiments are conducted on a Linux server with a GPU (NVIDIA RTX A40, Memory 40G). The training times of TGL-LLM are approximately 10 hours, while this may vary \wrt the size of the dataset and candidate set.

\begin{table*}
\centering
\caption{The results of TGL-LLM compared with Non-LLM methods and LLM-based methods. Bold and underlined indicate the best and the second-best performance, respectively.}
\vspace{-3mm}
\label{tab:overall_performance}
\renewcommand{\arraystretch}{1.3}
\resizebox{\linewidth}{!}{
\begin{tabular}{l|l|l|ccc | ccc |ccc }
\toprule
\multirow{2}{*}{\textbf{Model Type}} & \multirow{2}{*}{\textbf{Model}} & 
\multirow{2}{*}{\textbf{Input}} & \multicolumn{3}{c|}{\textbf{POLECAT-IR}} & \multicolumn{3}{c|}{\textbf{POLECAT-IS}} & \multicolumn{3}{c}{\textbf{POLECAT-EG}} \\
 & & & \textbf{Acc@4} & \textbf{Acc@6} & \textbf{Acc@10} & \textbf{Acc@4} & \textbf{Acc@6} & \textbf{Acc@10} & \textbf{Acc@4} & \textbf{Acc@6} & \textbf{Acc@10}\\ 

\midrule
\multirow{6}{*}{Non-LLM} & 
 \textbf{DistMult~\cite{DistMult}} & \multirow{6}{*}{Graph}
& 0.4901 & 0.4110 & 0.3291 
& 0.5364 & 0.4718 & 0.4060
& 0.4100 & 0.3339 & 0.2442\\
& \textbf{ConvTransE~\cite{ConvTransE}} & 
& 0.5169 & 0.4371 & 0.3563
& 0.5460 & 0.4846 & 0.4162
& 0.4562 & 0.3757 & 0.2736\\
& \textbf{RGCN~\cite{RGCN}} & 
& 0.5125 & 0.4372 & 0.3661
& 0.5355 & 0.4731 & 0.4071
& 0.4318 & 0.3499 & 0.2558\\
& \textbf{RENET~\cite{RENET}} 
&   
& 0.5099 & 0.4274 & 0.3448 
& 0.5399 & 0.4773 & 0.4077
& 0.4368 & 0.3539 & 0.2681
\\

& \textbf{REGCN~\cite{REGCN}} 
&  
& 0.4979 & 0.4264 & 0.3632
& 0.5353 & 0.4811 & 0.4227
& 0.4407 & 0.3568 & 0.2602\\

& \textbf{HisMatch~\cite{HiSMatch}} 
& 
& 0.5291 & 0.4562 & 0.3841
& 0.5478 & 0.4887 & 0.4260
& 0.4344 & 0.3529 & 0.2647\\

\midrule
\multirow{2}{*}{Zero-shot} 
& \textbf{GPT-3.5-turbo~\cite{OpenAI2022ChatGPT}} 
& \multirow{2}{*}{Text}  
& 0.4447 & 0.3396 & 0.2409
& 0.4627 & 0.3624 & 0.2480
& 0.3893 & 0.2917 & 0.1873 \\
& \textbf{GPT-4o-mini~\cite{achiam2023gpt}} 
& 
& 0.4907 & 0.3972 & 0.3090
& 0.4874 & 0.3960 & 0.3107
& 0.4082 & 0.3130 & 0.2093\\

\midrule
\multirow{4}{*}{Fine-tune}
& \textbf{GenTKG~\cite{GENTKG}} 
& \multirow{2}{*}{Text} 
& 0.5749 & 0.4806 & 0.3710
& 0.5863 & 0.4877 & 0.4129
& 0.4869 & 0.3911 & 0.2920 \\
& \textbf{CoH~\cite{Chain-of-History}}
&
& 0.6037 & 0.5074 & 0.4060 
& 0.6049 & 0.5158 & 0.4213
& 0.5092 & 0.4174 & 0.3153\\ 
\cmidrule{2-12}
& \textbf{KoPA~\cite{KoPA}} 
& \multirow{2}{*}{Graph} 
& \underline{0.6083} & \underline{0.5083} & \underline{0.4358}
& \underline{0.6110} & \underline{0.5261} & \underline{0.4336}
& \underline{0.5315} & \underline{0.4397} & \underline{0.3290} \\ 
& {\cellcolor[rgb]{0.902,0.902,0.902}}\textbf{TGL-LLM(ours)} 
& 
&
{\cellcolor[rgb]{0.902,0.902,0.902}}\textbf{0.8514} &{\cellcolor[rgb]{0.902,0.902,0.902}}\textbf{0.8269} &{\cellcolor[rgb]{0.902,0.902,0.902}}\textbf{0.7407}
&{\cellcolor[rgb]{0.902,0.902,0.902}}\textbf{0.8779} &{\cellcolor[rgb]{0.902,0.902,0.902}}\textbf{0.8459} &{\cellcolor[rgb]{0.902,0.902,0.902}}\textbf{0.7753}
&{\cellcolor[rgb]{0.902,0.902,0.902}}\textbf{0.8109} &{\cellcolor[rgb]{0.902,0.902,0.902}}\textbf{0.7697} &{\cellcolor[rgb]{0.902,0.902,0.902}}\textbf{0.6810} \\

\bottomrule
\end{tabular}}
\vspace{-2mm}
\end{table*}

\subsection{Performance Comparison (RQ1)}
We compare the overall performance of our proposed method, TGL-LLM, with several state-of-the-art methods for temporal knowledge graph forecasting. Table~\ref{tab:overall_performance} presents the overall performance of our method alongside all the baselines. From these results, we can draw the following observations. First, as we expected, our method outperforms all the baselines of both Non-LLM and LLM-based TKGF methods by a large margin across three datasets. We further adopt ANOVA significance test~\cite{anova} over the performances of all the baselines. All of the evaluated p-values are below 0.01, demonstrating the superior effectiveness of our method. Second, among the three types of baselines, LLM-based TKGF methods on the fine-tune setting generally perform better than the closed commercial LLMs and the Non-LLM method, indicating that supervised prompt tuning is crucial for enhancing the ability of LLMs to understand the task of TKGF. Furthermore, the embedding-based methods achieve better performance than the text-based methods across three datasets. We attribute this to the sufficient structural information provided by the graph embedding. Third, analyzing the results of Non-LLM methods, we observe that HisMatch performs best on the POLECAT-IR and POLECAT-IS datasets. However, for the POLECAT-EG dataset, the methods based on TKGs underperform compared to ConvTransE. The possible reasons are from two aspects: 1) insufficient modeling of temporal patterns, as evidenced by the poorer performance of REGCN compared to RGCN on POLECAT-IR and POLECAT-IS, and 2) smaller size of the POLECAT-EG dataset, particularly in the test set, which offers less useful temporal information.

\subsection{Ablation Study (RQ2)}
\subsubsection{The impact of hybrid graph tokenization}
\begin{table}
\centering
\caption{Performance comparison \wrt adopting different graph representations as graph tokens.}
\vspace{-3mm}
\renewcommand{\arraystretch}{1.2}
\label{tab:hybrid graph tokenization}
\resizebox{\linewidth}{!}{
\begin{tabular}{l| cc | cc |cc}
\toprule
 \multirow{2}{*}{\textbf{Model}} & \multicolumn{2}{c|}{\textbf{POLECAT-IR}}        & \multicolumn{2}{c|}{\textbf{POLECAT-IS}} & \multicolumn{2}{c}{\textbf{POLECAT-EG}}\\
 & \textbf{Acc@4} & \textbf{Acc@10} & \textbf{Acc@4} & \textbf{Acc@10} & \textbf{Acc@4} & \textbf{Acc@10}\\
\midrule
\textbf{Raw}  
& 0.5688 & 0.3510
& 0.5898 & 0.3950
& 0.4969 & 0.2993\\

\textbf{Static} 
& 0.6165 & 0.4084  
& 0.6106 & 0.4329
& 0.5278 & 0.3308 \\

\textbf{GRU}  
& 0.7098 & 0.6404
& 0.8528 & 0.7426
& 0.6674 & 0.5168 \\

\textbf{ConvTransE} 
& 0.8080 & 0.6230 
& 0.7841 & 0,6033
& 0.5039 & 0.3119\\

\midrule
\textbf{TGL-LLM} 
&\textbf{0.8514} &\textbf{0.7407} 
&\textbf{0.8779} &\textbf{0.7753}
&\textbf{0.8109} &\textbf{0.6810}\\
\bottomrule
\end{tabular}}
\vspace{-3mm}
\end{table}
To shed light on how insufficient modeling of temporal patterns affects the performance of LLMs, we compare four model variants: \textbf{Raw}(\ie without graph representation), \textbf{Static} (\ie with static graph embedding), \textbf{GRU} (\ie with the output of GRU), and \textbf{ConvTransE} (\ie with the output of ConvTransE). The results are presented in Table~\ref{tab:hybrid graph tokenization}. Clearly, \textbf{TGL-LLM} achieves the best performance by utilizing recent historical graph embeddings as graph tokens, demonstrating the sufficient modeling of temporal patterns in our proposed method. Next, we observe that \textbf{Static} underperforms compared to the graph representation with temporal information. This indicates that there are significant temporal patterns within our datasets, and effectively capturing these temporal patterns is crucial for LLMs to make accurate forecasts. Additionally, By comparing the performance of \textbf{GRU}, \textbf{CovTransE}, and \textbf{TGL-LLM}, we find that deeper graph representations lead to a decline in the performance of LLMs. This decline may be attributed to the fact that the deeper graph representations in the pre-trained graph models are challenging for the LLMs to comprehend and struggle to align effectively with the token embedding space.

\subsubsection{The impact of two-stage training paradigm}
\begin{table}
\centering
\caption{Performance comparison \wrt applying different sampling strategy.}
\renewcommand{\arraystretch}{1.2}
\label{tab:sample_strategy}
\resizebox{\linewidth}{!}{
\begin{tabular}{l| cc | cc |cc}
\toprule
 \multirow{2}{*}{\textbf{Model}} & \multicolumn{2}{c|}{\textbf{POLECAT-IR}}        & \multicolumn{2}{c|}{\textbf{POLECAT-IS}} & \multicolumn{2}{c}{\textbf{POLECAT-EG}}\\
 & \textbf{Acc@4} & \textbf{Acc@10} & \textbf{Acc@4} & \textbf{Acc@10} & \textbf{Acc@4} & \textbf{Acc@10}\\
\midrule
\textbf{Random}  
& 0.8467 & 0.6745
& 0.8631 & 0.7381 
& 0.7823 & 0.6511\\
\textbf{w/o-IF}  
& 0.8369 & 0.6871 
& 0.8622 & 0.7388
& 0.7920 & 0.6605\\

\textbf{w/o-DS} 
& 0.8354 & 0.6864 
& 0.8765 & 0.7731 
& 0.7673 & 0.6220\\
\midrule
\textbf{TGL-LLM} 
&\textbf{0.8514} &\textbf{0.7407} 
&\textbf{0.8779} &\textbf{0.7753}
&\textbf{0.8109} &\textbf{0.6810}\\
\bottomrule
\end{tabular}}
\vspace{-3mm}
\end{table}
We conduct experiments to evaluate the impact of the two-stage training paradigm for TKGF. Here, we showcase three variants to analyze the effect: \textbf{Random} (\ie fine-tune on the randomly sampled training subset), \textbf{w/o-IF} (\ie replace the influence function with the random sampling), and \textbf{w/o-DS} (\ie without diversity sampling). As illustrated in Table ~\ref{tab:sample_strategy}, it is clear that the two-stage training paradigm outperforms the other three variants, which contributes to the effective cross-modal alignment between graph and language. Moreover, we find that the absence of either the high-quality subset or the diversity subset can lead to worse performance compared to random sampling, demonstrating that high-quality and diverse data are equally important for LLMs. It is worth pointing out that the performance on the POLECAT-IS dataset remains strong without diversity sampling. This may be because the temporal and relational patterns in the testing set are more similar to those in the training set for POLECAT-IS.

\subsection{Model Study (RQ3)}
We conduct more experiments to study some key properties of our model.
\subsubsection{Long-tail Issue}
\begin{figure}
    \centering
    \includegraphics[width=\linewidth]{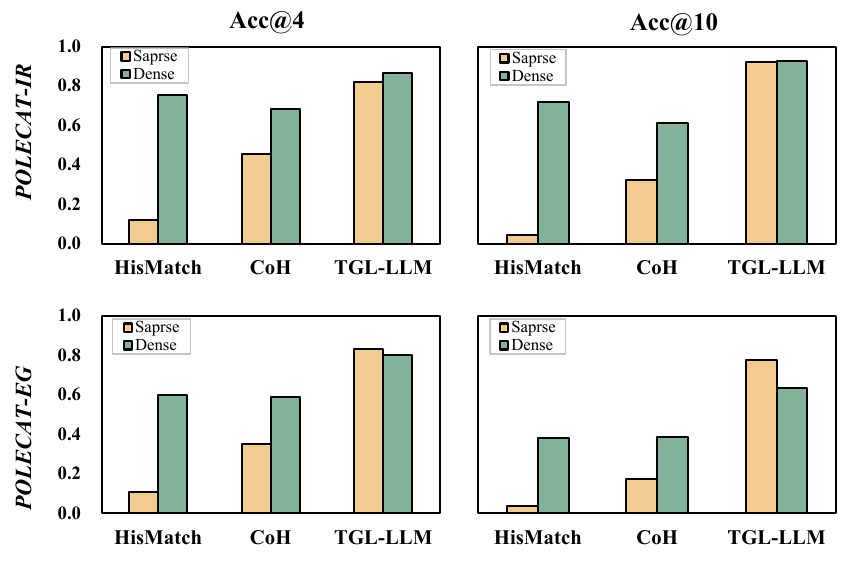}
    \caption{The performance comparison across two datasets on the long-tail entities.}
    \label{fig: long-tail}
    \vspace{-3mm}
\end{figure}
Several previous studies~\cite{SeCoGD,evaluati-tef} have identified that traditional methods still struggle with the prevalent issue of long-tail entities. To evaluate the performance of TGL-LLM on long-tail entities in TKGs, we re-split the test dataset into two subsets: dense and sparse. The dense subset includes queries where the query object appears over 100 times in the training dataset, while the remaining facts form the sparse subset. For instance, in the POLECAT-IR dataset, the dense subset has 353 objects and 7,099 facts, while the sparse subset contains 1,023 objects and 3,954 facts. The results across two datasets are shown in Figure~\ref{fig: long-tail}. Benefiting from the open-world knowledge in LLMs, the LLM-based methods achieve better performance than Non-LLM methods. In contrast, with sufficient information from the training dataset, Non-LLM methods remain competitive. Notably, under no matter what specific settings, TGL-LLM demonstrates the strongest overall performance and has much smaller gaps, indicating that LLM can still effectively tackle the task of TKGF on long-tail entities with extensive knowledge and effective alignment between graph and language. 

\subsubsection{The impact of historical length}

\begin{figure}
    \centering
    \includegraphics[width=0.75\linewidth]{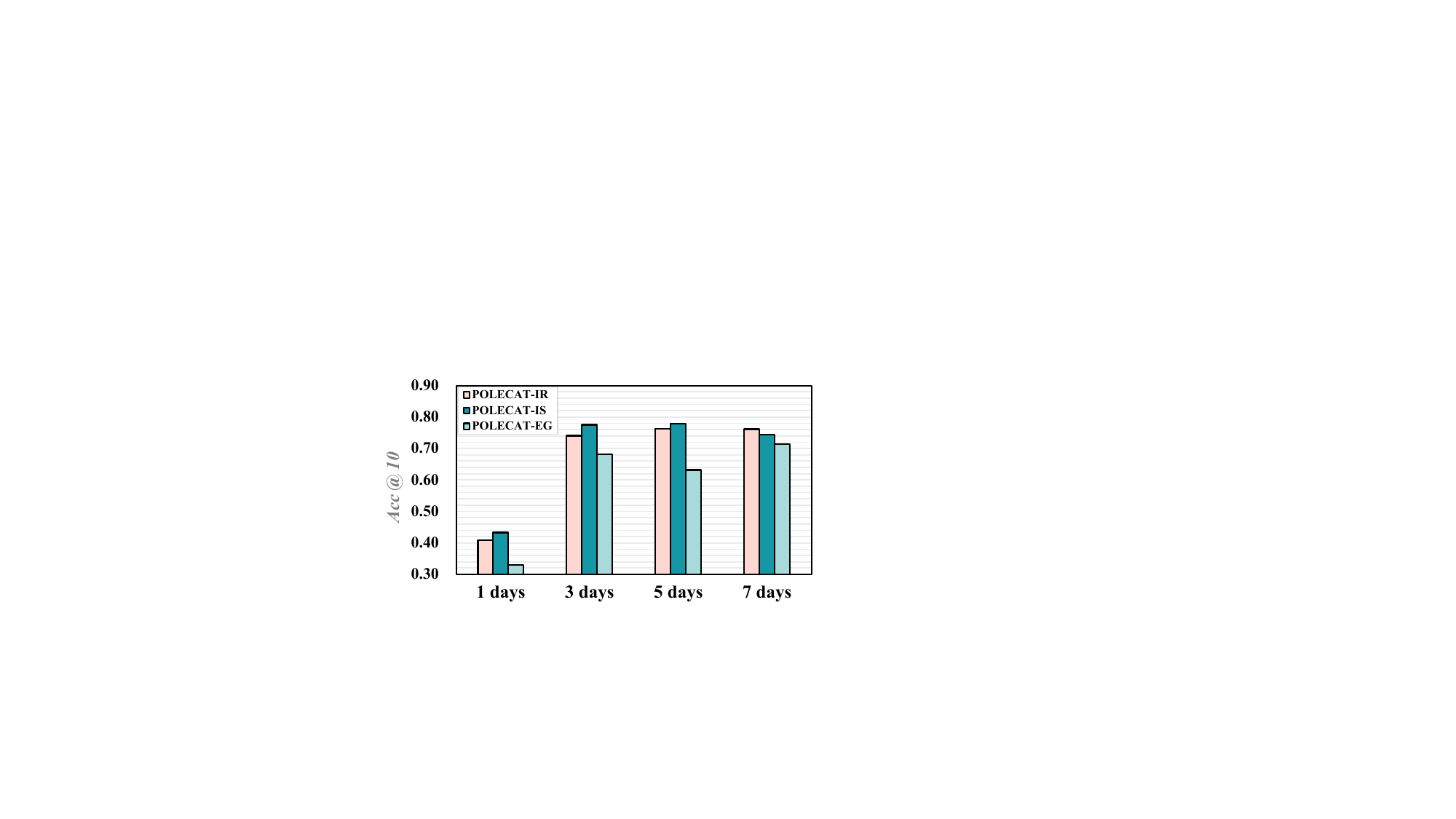}
    \vspace{-3mm}
    \caption{Performance comparison of our method when using varying history length.}
    \label{fig: historical_length}
\end{figure}
The scope of history plays a crucial role in effective forecasting. Driven by this hypothesis, we vary the historical length of graph embedding, and the result is shown in Figure~\ref{fig: historical_length}, which illustrates that 5 is the best for POLECAT-IR and POLECAT-IS and 7 is the best for POLECAT-EG. Furthermore, we find that as the historical length increases, the performance starts to converge, which indicates that the expansion of history length comes along with more noisy information.

\subsubsection{Training Costs}
In order to evaluate the training costs of various LLM-based methods, we conduct experiments on POLECAT-IR, as shown in Table~\ref{tab: training cost}. \textbf{Avg.Token}, \textbf{GPU (GiB)}, and \textbf{Time(h)} represent the average token length of various LLM-based methods, the GPU memory usage, and the training time required for fine-tuning the LLMs, respectively. Compared to the text-based method, the training cost of the embedding-based method significantly decreases due to the integration of graph embedding. Although KoPA has the lowest training cost, TGL-LLM incurs a slightly higher training cost while endowing  LLMs with enhanced reasoning capabilities over TKGs.
\vspace{-3mm}
\begin{table}
\centering
\vspace{-3mm}
\caption{The comparison of training costs \wrt applying different LLM-based methods. The statistics are based on NVIDIA RTX A40.}
\vspace{-3mm}
\renewcommand{\arraystretch}{1.2}
\label{tab: training cost}
\resizebox{\linewidth}{!}{
\begin{tabular}{l| ccc |c}
\toprule
\textbf{Model} & \textbf{Avg.Token} & \textbf{GPU (GiB)} & \textbf{Time (h)} & \textbf{Acc@10} \\
\midrule
\textbf{CoH}
& 466 & 36.78 & 21.53 & 0.406 \\
\textbf{KoPA}
& \textbf{223} & \textbf{22.98} & \textbf{10.23} & 0.4358 \\
\textbf{TGL-LLM}
& 245 & 23.08 & 11.26 & \textbf{0.7407} \\
\bottomrule
\end{tabular}}
\vspace{-3mm}
\end{table}

\section{Conclusion and Future Work}
In this paper, we integrated temporal graph learning into LLM-based temporal knowledge graph model. We introduced a new framework, TGL-LLM, which consists of two critical components: hybrid graph tokenization and a two-stage training paradigm to address two major limitations of existing embedding-based methods: the insufficient modeling of temporal patterns and the ineffective cross-modal alignment between graph and language respectively. 
Empirical results demonstrated that TGL-LLM significantly outperforms all baselines in TKGF, underscoring its effectiveness and superior performance.

Moving forward, we will shift our focus to: 1) developing more effective graph adapters instead of relying solely on a simple multi-layer perceptron (MLP); 2) endowing LLMs with capabilities for generative forecasting; 3) not limiting to global graph representations, but rather introducing more local graph representations to obtain more accurate contextual information.; and 4) improving the performance of LLMs across a broader range of subtasks, including relation prediction, time interval prediction, long-horizon prediction, and event distribution prediction, among others. And we hope that TGL-LLM can inspire the researchers to explore more powerful temporal patterns modeling within LLMs, paving the way to the development of temporal knowledge graph forecasting.

\bibliographystyle{ACM-Reference-Format}
\bibliography{main}

\appendix
\section{Appendix}
\subsection{Details on Dataset Construction}
\label{dataset_constr}

This section presents the sources and steps involved in constructing a new benchmark dataset for event prediction using large language models. To address the issue of information leakage in event prediction methods using LLMs—where the pre-training data of LLMs may already include events from the benchmark test set, making evaluation less reliable—we construct a new benchmark dataset based on POLECAT, a publicly available event dataset. POLECAT contains events extracted from internet news sources starting in 2018, including data from after the LLM's training cutoff date, which can be used for evaluation. The construction pipeline consists of the following steps.

\paragraph{Filter by Country and Time} To reduce the size of the POLECAT, we applied filters based on both time and country. We downloaded data from January 1, 2018, to April 21, 2024, focusing on events related to Iran, Israel, and Egypt. The country filtering was applied to three fields: Actor Country, Recipient Country, and Country. Any event where at least one of these fields contained Iran, Israel, or Egypt was retained.

\paragraph{Remove Incomplete Events} Since POLECAT is extracted from news articles using event extraction algorithms, there are cases where the extraction process fails or results in missing information. To improve data quality, we removed events where both Actor Name and Actor Name Raw were listed as 'None', `np.NaN`, or empty strings. Actor Name Raw represents the original entity text as it appeared in the news sentence, while Actor Name is the result after entity mapping. When both fields are missing, it indicates a failure in the extraction process, with no fallback to the original entity name from the text. Similarly, we excluded events where both Recipient Name and Recipient Name Raw were missing.

\paragraph{Entity Expansion} The event extraction algorithm used by POLECAT often extracts multiple actors and recipients for a single event. Specifically, the Actor Name, Actor Name Raw, Recipient Name, and Recipient Name Raw fields may contain multiple entities within one data point, separated by semicolons. To maintain consistency with the previous steps in the dataset processing, we split these rows. For events with multiple actors and/or recipients, we create separate rows for each possible combination of actors and recipients. In this way, each unique actor-recipient pair is represented in its own row, while the event details, such as event type and other metadata, remain unchanged. If any Actor Name or Recipient Name is missing after expansion, we follow the same logic as before by filling it with the corresponding "Raw" fields or removing the row if no replacement is available.

\paragraph{Duplicate Removal} To eliminate redundant data, duplicates are identified based on five key fields: Actor Name, Recipient Name, Event Type, Event Mode, and Event Date. We removed any such duplicated entries to ensure that each unique event is represented only once in the dataset.

\begin{figure}[h]
    \centering
    \includegraphics[width=\linewidth]{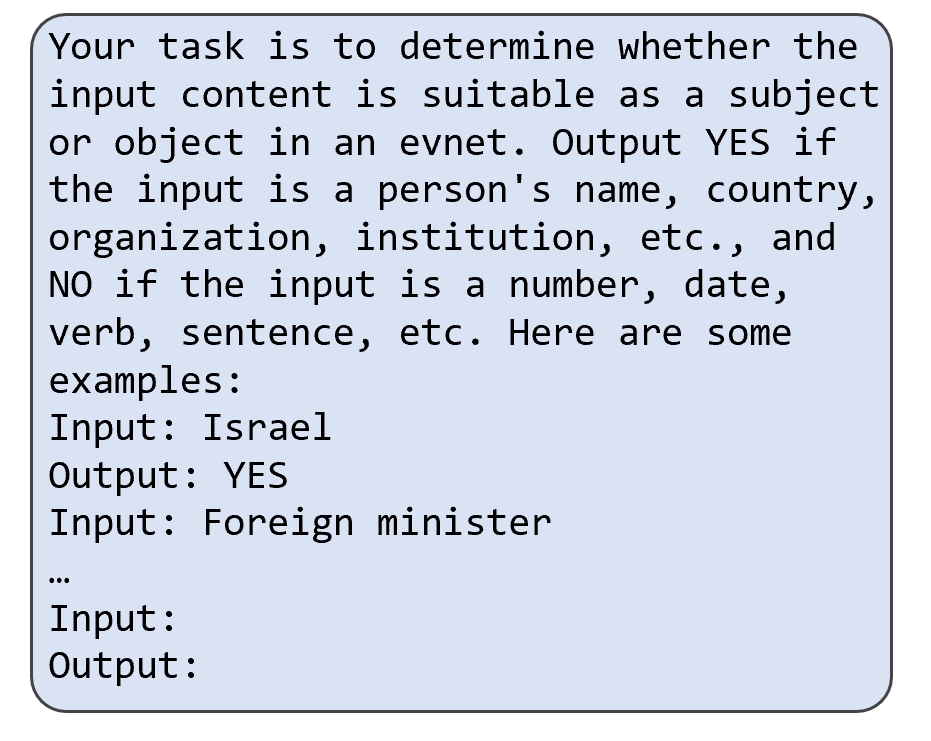}
    \caption{Specific prompt used for entity filtering.}
    \label{fig:dataset_statistics}
    \vspace{-3mm}
\end{figure}

\paragraph{Entity Filtering} To improve the quality of the dataset, we removed irrelevant or incorrectly extracted entities, which may have resulted from algorithmic errors or noise. Using GPT-3.5-Turbo, we evaluated each unique entity in the dataset. The model was prompted to assess whether an entity was meaningful or not, responding with "yes" or "no." Entities marked "yes" were kept. For entities marked "no," we checked their frequency. If they appeared in 5 or fewer events, they were removed; otherwise, they were retained. The specific prompt used is shown in Figure \ref{fig:dataset_statistics}.

\subsection{Dataset Statistics}

\label{additional_stats}

\paragraph{Overview} Our new dataset, constructed on the basis of POLECAT from filtered events extracted from news articles, spans from January 2018 to April 2024.

\begin{figure}
    \centering
    \includegraphics[width=\linewidth]{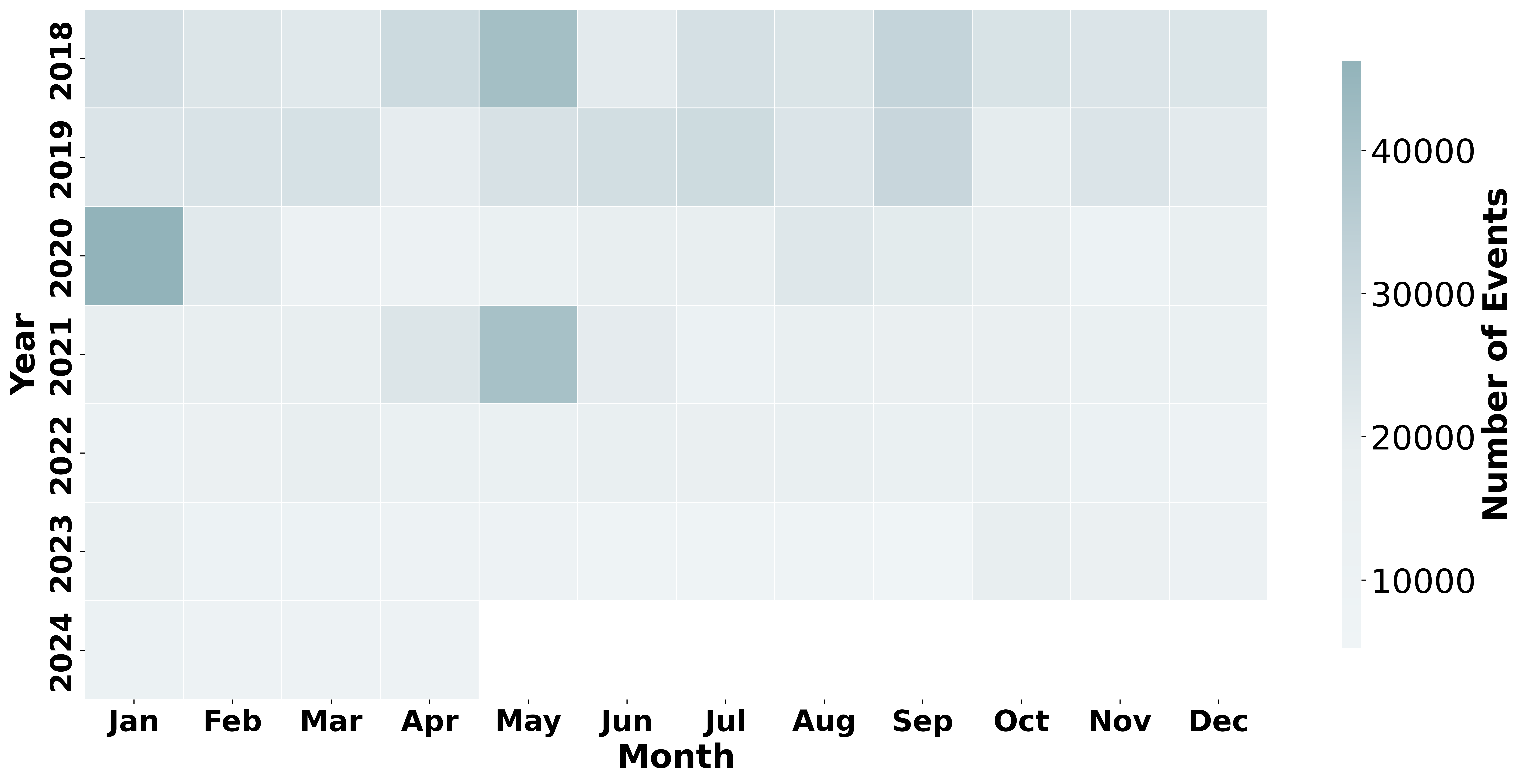}
    
    \caption{Heatmap illustration of event distribution by year and month.}
    \label{fig:event_distribution}
    \vspace{-3mm}
\end{figure}

\begin{figure}
    \centering
    \includegraphics[width=\linewidth]{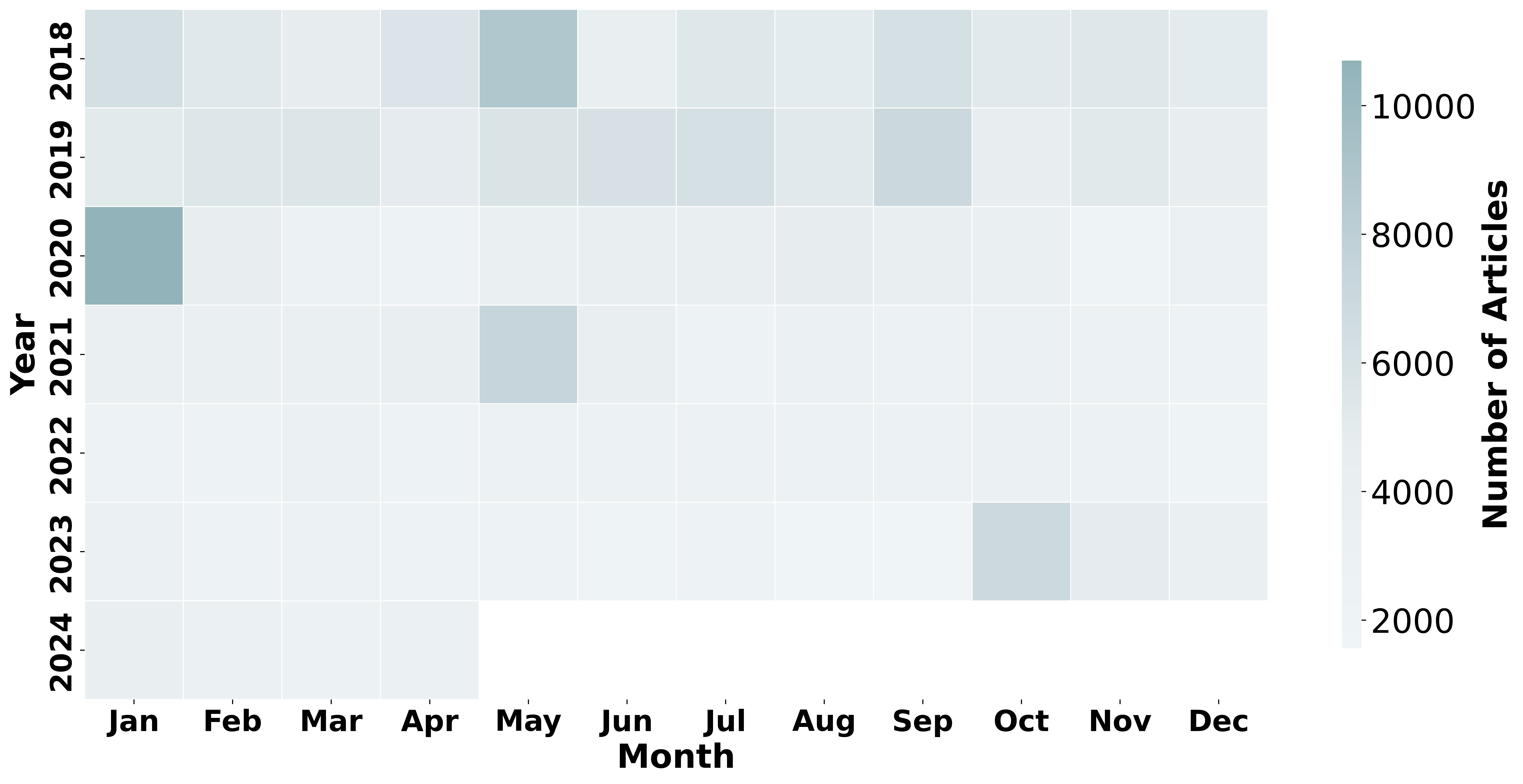}
    \caption{Heatmap illustration of article distribution by year and month.}
    \label{fig:article_distribution}
    \vspace{-3mm}
\end{figure}

\paragraph{Time Distribution} To provide insights into the temporal distribution of the dataset, we analyzed the Event Date field to examine how the data is distributed over time. We drew two heatmaps: one showing the distribution of events by year and month, and the other showing the distribution of news articles by year and month. One news article may contain multiple events, providing additional context to the analysis. These visualizations help to identify temporal patterns, such as peaks or trends in event occurrences (see Figures~\ref{fig:event_distribution} and~\ref{fig:article_distribution}).

\begin{figure}
    \centering
    \includegraphics[width=\linewidth]{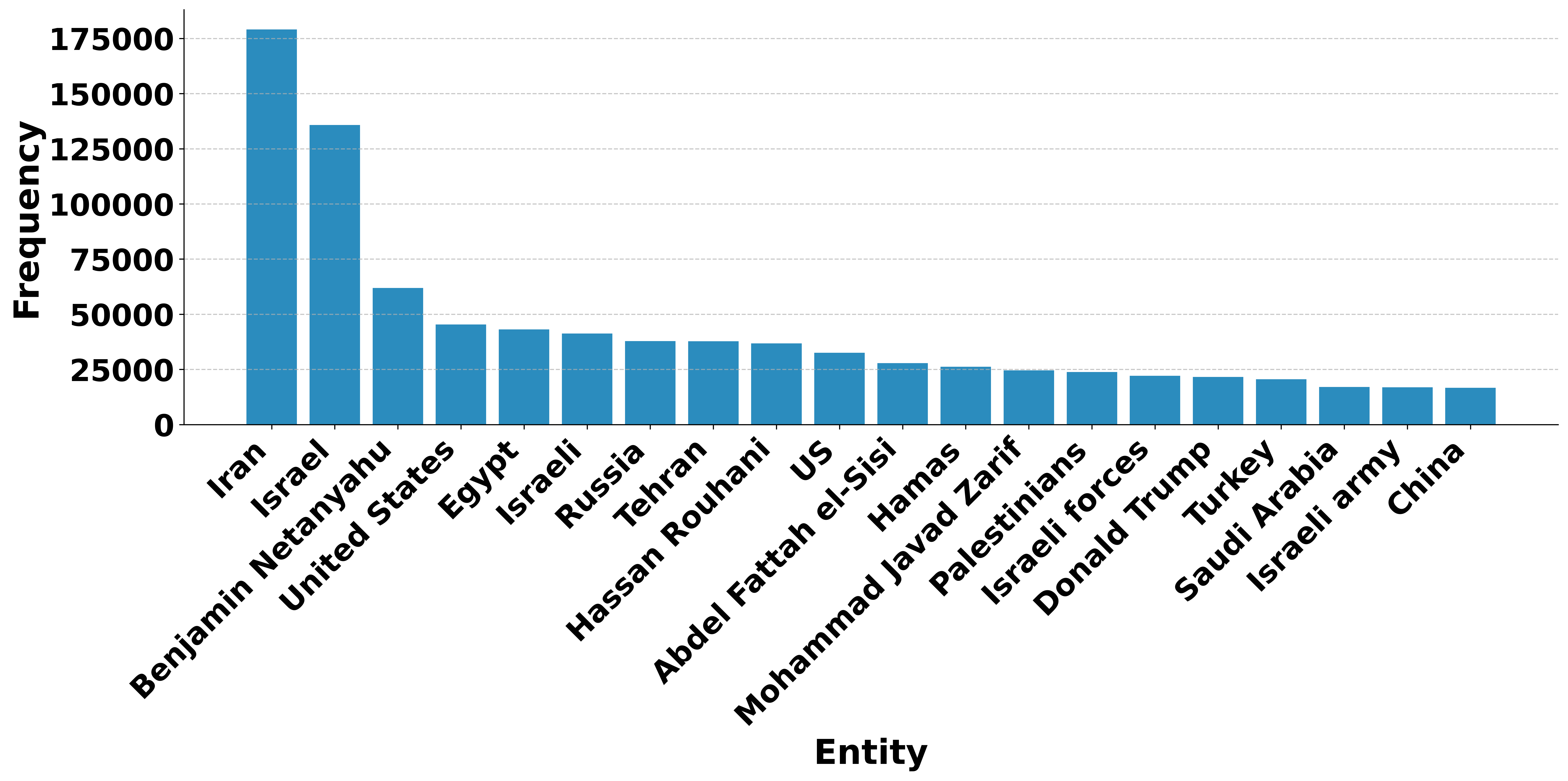}
    \caption{The top-20 most frequent entities in the dataset.}
    \label{fig:entity_distribution}
    \vspace{-3mm}
\end{figure}

\paragraph{Entity Distribution} In this section, we analyze the distribution of entities within the dataset. Entities are collected from two key fields: \textit{Actor Name} and \textit{Recipient Name}, representing the primary and secondary participants in each event, respectively. We ranked the top 20 most frequently occurring entities across the dataset. These entities represent the individuals, organizations, or groups that are most involved in or affected by the events in Iran, Israel, and Egypt during the given time period. The frequency of occurrence for each entity was calculated, and we visualized this distribution with a bar chart (see Figure~\ref{fig:entity_distribution}).

\begin{figure}
    \centering
    \includegraphics[width=\linewidth]{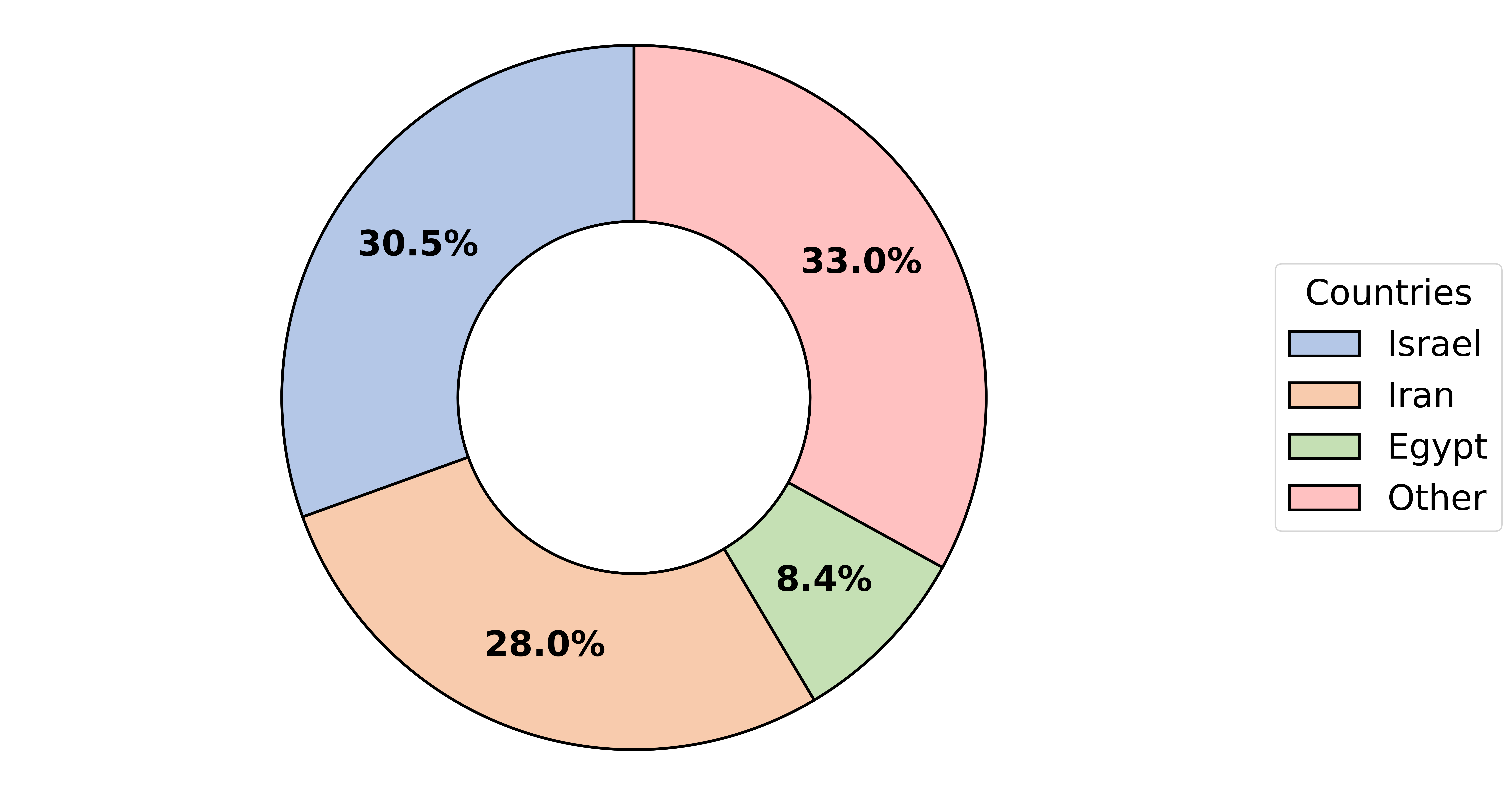}
    \caption{The event distribution across countries.}
    \label{fig:country_distribution}
    \vspace{-3mm}
\end{figure}

\paragraph{Country Distribution} We analyzed the distribution of events related to Egypt, Iran, and Israel. Entities from these countries were identified from the \textit{Actor Name} and \textit{Recipient Name} fields. We counted the frequency of events for each country. A bar chart was generated to display the number of events involving Egypt, Iran, and Israel. Although we applied filtering, entity expansion during the process introduced other related entities, which are still relevant to the events (see Figure~\ref{fig:country_distribution}).

\subsection{Statistical Analysis of Generative vs. MCQ Settings.}
\label{statistic}

In this section, we explore the differences between the generative and MCQ settings for temporal knowledge graph forecasting tasks using LLM. We include this statistical analysis to demonstrate that the MCQ and generative settings mentioned in Section~\ref{preliminary} are fundamentally equivalent when LLM is applied to TKG learning. 

We conducted an empirical analysis using the GenTKG model, which uses rules to retrieve historical context for prompt construction in a generative setting, and then requires the LLM to make predictions using the generative setting instead of MCQ. We aim to demonstrate that historical information serves as an implicit MCQ, by analyzing the proportion of entities generated by the LLM that appear in the historical information provided in the prompt. This would show that the generative setting is essentially no different from the MCQ setting. We conducted experiments on two datasets, ICEWS and GDELT. The results are shown in Table \ref{tab:gtkg_context_stats}.

\begin{table}[h]
    \centering
    \caption{Statistics of Ground-truth (GT) and GenTKG Predictions from Context (Ctx). 
    GT refers to ground-truth entities. Gen. Entities from Ctx refers to the percentage of generated entities that appear in the context.}
    \renewcommand{\arraystretch}{1.2}
    \resizebox{0.75\linewidth}{!}{
    \begin{tabular}{l|c|c}
        \toprule
        \multicolumn{1}{l|}{\textbf{Statistic}} & \textbf{ICEWS} & \textbf{GDELT} \\
        \midrule
        \multicolumn{1}{l|}{\textbf{Test Num}} & 7,371 & 9,715 \\
        \multicolumn{1}{l|}{\textbf{GT in Context (\%)}} & 63.60 & 45.58 \\
        \multicolumn{1}{l|}{\textbf{Gen. Entities from Ctx (\%)}} & 96.97 & 94.85 \\
        \midrule
        \multicolumn{1}{l|}{ \textbf{Correct (Ctx, \%)}} & 37.59 & 11.83 \\
        \multicolumn{1}{l|}{\textbf{Correct (Non-Ctx, \%)}} & 0.01 & 0.03 \\
        \multicolumn{1}{l|}{\textbf{Wrong (\%)}} & 62.39 & 88.13 \\
        \bottomrule
    \end{tabular}}
    \label{tab:gtkg_context_stats}
\end{table}

Based on the results, we can see that in both datasets, the majority of entities generated by the LLM were already present in the prompt, indicating that the LLM rarely generates entirely new entities. This suggests that the LLM, when making predictions, tends to select entities from the provided context rather than truly generating new ones, meaning that the generative setting does not lead to genuine generation. Furthermore, the performance on the ICEWS dataset was significantly better than on GDELT. This could be because, during context prompt generation, the proportion of correct answers included in the context prompts was notably higher in ICEWS compared to GDELT. This indicates that even in the generative setting, the model's performance is heavily influenced by the context prompt, making it no different from the MCQ setting.

Therefore, pure generation without any given context remains extremely challenging for large language models. Simply changing the problem setting does not result in genuine generation.

\subsection{Baseline Implementation Details}

This section provides a detailed explanation of the baseline mentioned in the Baseline section.

Non-LLM Methods including:
\begin{itemize}
    \item DistMult~\cite{DistMult} uses learned relation representations to model logical rules, primarily focusing on representing entity relationships in a simple multiplicative form.
    
    \item ConvTransE~\cite{ConvTransE} applies convolution to concatenated entity and relation embeddings, helping to model more complex relationships and enhance link prediction capabilities.
    
    \item RGCN~\cite{RGCN} employs relation-specific transformations with weight matrices for each relation. To handle many relations and prevent overfitting, it uses basis decomposition and block diagonal decomposition.
    
    \item RENET~\cite{RENET} builds on sequential and graph-based modeling to capture both temporal and structural interactions. It explicitly incorporates temporal sequences for dynamic knowledge graphs.
    
    \item REGCN~\cite{REGCN} incorporates a recurrent unit to capture temporal dependencies in addition to structural information. It also uses a static graph to improve initial embeddings.
    
    \item HisMatch~\cite{HiSMatch} reformulates temporal knowledge graph forecasting into a query-candidate matching problem, using a two-branch framework to efficiently match queries to candidate entities.
\end{itemize}

Zero-Shot Methods use TKG information in text form in the LLMs prompt. In our implementation, we used the approach described in the CoH~\cite{Chain-of-History}) paper to construct text prompts and accessed GPT-3.5-turbo~\cite{OpenAI2022ChatGPT} and GPT-4o-mini~\cite{achiam2023gpt} through the OpenAI API to obtain answers.

For fine-tune methods, we implement:

\begin{itemize}
    \item CoH~\cite{Chain-of-History} uses schema matching, entity-augmented, and relation-augmented historical facts to construct text prompts for fine-tuning LLMs.
    \item GenTKG~\cite{GENTKG} learns temporal logic rules to construct text prompts for fine-tuning LLMs.
    \item KoPA~\cite{KoPA} employs structural embedding pre-training to capture knowledge graph (KG) information and transforms it into textual prefixes to guide LLMs.
\end{itemize}

\subsection{Instruction}
\label{instruction}
\begin{figure}
    \centering
    \includegraphics[width=0.75\linewidth]{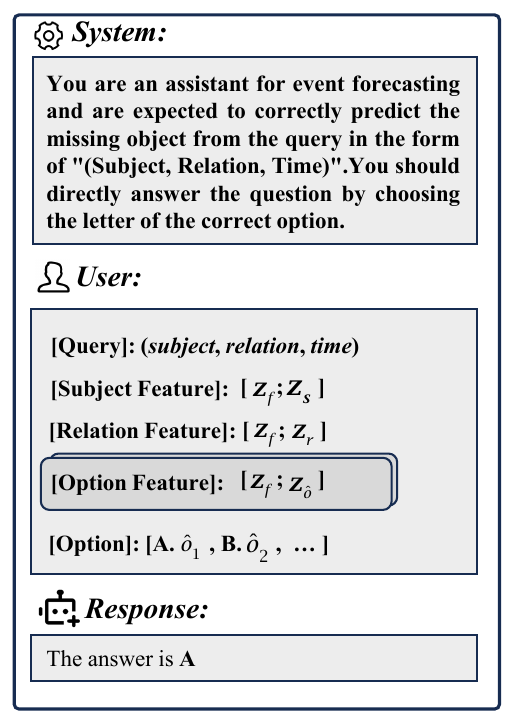}
    \caption{The instruction utilized in the hybrid graph tokenization.}
    \label{fig:instruction}
    \vspace{-3mm}
\end{figure}

As illustrated in Figure~\ref{fig:instruction}, we present the instruction used for hybrid graph tokenization, which includes the textual task definition, as well as the format of the query and input.

\end{document}